\documentstyle[11pt,times,epsf,psfrag,dsfont,wrapfig,graphics,graphicx]{article}

\newcommand{\NN}{{\mathds{N}}}
\newtheorem{obs}{Observation}[section]

\newenvironment{proof}{\par\noindent{\bf Proof:}}{\mbox{}\hfill$\Box$\\}

\newtheorem{definition}{Definition}[section]
\newtheorem{theorem}{Theorem}[section]

\newtheorem{lemma}{Lemma}[section]

\newtheorem{corollary}{Corollary}[section]
\epsfverbosetrue

\begin{document}
\title{Faster Streaming Algorithms for Graph Spanners}

\author{Surender Baswana\\
        Department of Computer Science \& Engineering\\
        Indian Institute of Technology Kanpur\\
        Kanpur - 208016,~ INDIA.\\
        Email : {\texttt{sbaswana@cse.iitk.ac.in}}
}
\date{}
\maketitle
%
%
\begin{abstract}
Given an undirected graph $G=(V,E)$ on $n$ vertices, $m$ edges, and 
an integer $t\ge 1$, a subgraph $(V,E_S)$, $E_S\subseteq E$ is called a 
$t$-spanner if for any pair of vertices $u,v \in V$, the distance between 
them in the subgraph is at most $t$ times the actual distance.
We present streaming algorithms for computing a $t$-spanner of essentially
optimal size-stretch trade offs for any undirected graph. 

Our first algorithm is for the classical streaming model and works
for unweighted graphs only. The algorithm performs a single pass on the stream 
of edges and requires $O(m)$ time to process the entire stream of edges. 
This drastically improves the previous best single pass streaming algorithm for
computing a $t$-spanner which requires $\theta(mn^{\frac{2}{t}})$ time to 
process the stream and computes spanner with size slightly larger than the 
optimal.

Our second algorithm is for {\em StreamSort} model introduced by Aggarwal 
{\em et al.} \cite{ADRR:04}, which is the streaming model augmented with a 
sorting primitive. The {\em StreamSort} model has been shown to be a more 
powerful and still very realistic model than the streaming model
for massive data sets applications. Our algorithm, which works of weighted
graphs as well, performs $O(t)$ passes using $O(\log n)$ bits of working
memory only. 

Our both the algorithms require elementary data structures.

%
%
\end{abstract}

\noindent{\bf Keywords :} streaming, spanner, approximate shortest path
\pagebreak
%
\section{Introduction}\label{intro}
A spanner is a (sparse) subgraph  of a given graph that preserves approximate
distance between each pair of vertices. Putting in more formal words, a 
$t$-spanner of a graph $G=(V,E)$, for any $t\in \NN$ is a subgraph 
$(V,E_S), E_S\subseteq E$ such that, for any pair of vertices, 
their distance in the subgraph is at most 
$t$ times their distance in the original graph. 
The parameter $t$ is called the {\em stretch factor} 
associated with the $t$-spanner. The concept of spanners was defined formally 
by Peleg and Sch\"{a}ffer \cite{PS:89} though the associated notion was used
implicitly by Awerbuch \cite{Aw:85} in the context of network synchronizers.
Since then, spanner has found numerous applications in
the area of distributed systems, communication networks and all pairs
approximate shortest paths \cite{Aw:85,BS:04,PUl:87,PU:89}.

%
Each application of spanners requires, for a specified $t\in \NN$, 
a $t$-spanner of smallest possible size (the number of edges). 
Based on the famous girth conjecture by Erd\H{o}s \cite{Erdos:63}, 
Bollob\'{a}s \cite{Bol:78}, and Bondy and Simonovits \cite{BS:74}, it follows 
that for any $k\in \NN$, there are graphs on $n$ vertices 
whose $(2k-1)$-spanner or a $2k$-spanner 
will require $\Omega(n^{1+1/k})$ edges. The conjecture has been 
proved for $k=1,2,3$ and $5$. Note that the conjectured worst case lower bound is the same
for stretch $2k$ and $2k-1$, and by definition, a $(2k-1)$-spanner is also a $2k$-spanner. Therefore, from the perspective of an algorithmist,
the aim would be to design an efficient algorithm to compute a $(2k-1)$-spanner
whose size is $O(n^{1+1/k})$ for any given graph.

For unweighted graphs, Halperin and Zwick \cite{HZ:96} designed a deterministic
$O(m)$ time algorithm to compute a $(2k-1)$-spanner of $O(n^{1+1/k})$ size.
However, for weighted graphs, it took a series of improvements 
\cite{ADD:93,ABCP:98,Cohen:98,TZ:05,BS:03,BS-J:03} till an expected $O(m)$ 
time algorithm for computing a $(2k-1)$-spanner could be designed. 
This linear time randomized algorithm \cite{BS:03,BS-J:03} computes a 
$(2k-1)$-spanner of size $O(kn^{1+1/k})$ for a given weighted graph. 
Recently Roditty {\em et al.} \cite{RTZ:05} derandomized 
this algorithm.

%
%

In this paper, we consider the problem of computing a $(2k-1)$-spanner in
streaming model and its recently extended variant {\em StreamSort}. These 
models capture the complexities of algorithms designed for 
massive data set applications more accurately, and are thus gaining ever 
increasing attention these days. Our algorithms for computing spanners 
are significantly superior to the previously existing ones, and are arguably 
optimal. We shall now briefly describe the streaming model, the 
{\em StreamSort} model, and the motivation for computing spanners in streaming 
environment. Then we present the (bounds of) previously
existing streaming algorithms for spanners, and new results.

\subsection{Streaming model}
The streaming model \cite{HRR:99} has the following two characteristics : 
firstly the input data can be accessed sequentially (in the form of a stream), 
secondly the working memory is considerably smaller than the the size of the 
entire input stream. So an algorithm in this model can only make a few  
passes over the input stream to solve the corresponding problem. The sequentiality in accessing the data and the small working memory size enforce the following restriction : during a pass, a data item once evicted from the memory can't be brought back into the
working memory. It is due to this restriction that the streaming model is more 
stringent than other models namely, various external memory models 
\cite{AACS:87,VV:96}, and models for competitive analysis of algorithms
\cite{KMRS:88}.

The features and restrictions of the streaming 
model have been motivated by various technological factors pertaining to
massive data set applications.
%
%
%
%
%
Due to enormity of size along with various practical and economical reasons, the
input data of a massive data set application resides on secondary and tertiary 
storage devices. These devices are optimized for sequential access and impose 
substantial penalties (seek times, cache misses, pipeline stalls) for 
non-sequential data access. So an efficient algorithm in this model should
make a small number of sequential passes over the input data with a small size 
of working memory. The number of passes and the size of working memory are the 
two parameters associated with a streaming algorithm. An additional parameter
is the processing time per data item. These three parameters also capture
the efficiency criteria for a streaming algorithm.

This model
is gaining a lot of attention currently due to emerging massive data set 
applications. 
Earlier, in this model, much attention was 
given to problem related to computing order statistics, outliers, histograms
\cite{AMS:99,CCF:04,GGIKMS:02,HRR:99}. Recently, much attention has been given
to solving graph problems in this model, for example, approximate distances, 
spanners, and matching \cite{FKM+:04,FKM+:05,McG:05}. 
A typical graph problem in the streaming model involves making one or more 
sequential passes over the stream of edges. 

Aggarwal et al. \cite{ADRR:04} introduce an extension of streaming model called
{\em StreamSort} model which is more powerful and still very practical than 
the streaming model. An algorithm in this model performs two kinds of passes - 
{\em stream} pass and {\em sort} pass. The {\em stream} pass sequentially 
reads the input stream, processes it with its limited memory, and produces an 
output stream. During the pass, the output stream is written left to right, and
a data item once written can't be erased. A {\em sort} pass sorts a stream 
according to some well defined order and produces as an output a sorted stream.
An output stream of one pass can be used as input stream for the next
({\em stream} or {\em sort}) pass. An algorithm in {\em StreamSort} model thus 
performs a few stream pass and a few sort passes to solve a computational 
problem.
%
%
%
In a slightly simpler variant of {\em StreamSort} model, Demetrescu 
{\em et al.} \cite{DFR:06} presented streaming algorithms for undirected 
connectivity and shortest paths problem which achieve near optimal 
trading off between space and the number of passes.

\subsection{Computing a spanner in streaming environment and new results}
Being one of the fundamental problem in its own right, computing spanners
in a streaming environment is a significant problem. This problem has recently
gained more relevance due to all-pairs approximate shortest path problem in 
streaming environment. Due to enormity of size, it is just not feasible to 
compute or store all-pairs distances in streaming environment for graphs 
appearing in massive data sets applications. So one wants to 
settle for approximate shortest paths to save space. 
A result of Thorup and 
Zwick \cite{TZ:05} showed that for any data structure capable of answering
$(2k-1)$-approximate distance query would need $\Omega(n^{1+1/k})$ space.
The result obviously holds for streaming environment too. 
If we can compute $(2k-1)$-spanner efficiently in streaming environment,
it can be employed to solve the APASP problem in the streaming environment
in the following way : For any pair of vertices, just explore the spanner to
report the approximate distance. Feigenbaum {\em et al.} \cite{FKM+:05} took
this approach for all-pairs approximate distances in streaming environment.

We would like to add a note that a $k$-pass streaming algorithm 
for a $(2k-1)$-spanner of $O(kn^{1+1/k})$ size for any weighted graph is 
implicit in the algorithm of \cite{BS:03,BS-J:03},
and the processing time for each edge is also just $O(1)$
during each pass. The working memory required has size $O(kn^{1+1/k})$.
Since each pass is a time consuming task, it is always desirable to have a 
singe pass algorithm for computing a $(2k-1)$-spanner. For such an algorithm, 
one would also aim to keep processing time per edge bounded by a constant. 
Feigenbaum {\em et al.} \cite{FKM+:05} made a step in this direction. As a 
main result in their paper, they present a single pass streaming algorithm 
(Theorem 2.1, \cite{FKM+:05}) for computing a $t$-spanner for any unweighted 
graph. Though they don't mention it, their algorithm is indeed an adaptation of
the algorithm of \cite{BS:03,BS-J:03} for streaming environment. However, the 
bounds their algorithm achieves are suboptimal : For any 
$k\in \NN$, their algorithm computes a $(2k+1)$-spanner of expected size 
$O(kn^{1+1/k})$ and requires expected $\theta(k^2n^{1/k})$ processing time per 
edge. Note that the size of the spanner thus computed is away from the optimal 
by a factor of $\theta(n^{1/k^2})$. 


In this paper, we succeed in achieving optimal bounds and size-stretch trade 
offs for computing a $(2k-1)$-spanner in streaming environment. We achieve the 
following two results. 
\begin{enumerate}
\item
Given any unweighted undirected graph, and $k\in \NN$, a $(2k-1)$-spanner of
expected $O(kn^{1+1/k})$ size can be computed in classical streaming
model with single pass and $O(m)$ processing time for the entire stream 
(amortized constant processing time per edge). 

{\em Remark}.~ The algorithm at each stage maintains a 
$(2k-1)$-spanner of the graph seen so far. Therefore, it can also be viewed
as a partial dynamic (incremental) algorithm for computing a $(2k-1)$-spanner 
of an unweighted graph with amortized $O(1)$ time per edge insertion 
(the same observation, but with inferior bounds, holds for the earlier 
algorithm of \cite{FKM+:05}). \\

If the edges appear sorted in nondecreasing order of their weights in the
stream, our algorithm, without any modification at all, would work
for weighted graphs as well. As a result, it requires one {\em sort} pass 
followed by a {\em stream} pass  in the {\em StreamSort} model for computing a 
$(2k-1)$-spanner of expected $O(kn^{1+1/k})$ size for any $k\in \NN$ and
any weighted graph. Note that working memory has size of the order of 
spanner size, and though larger than $n$, is indeed optimal for classical 
streaming model.
\item
Given a weighted undirected graph, and $k\in \NN$, a $(2k-1)$-spanner
of expected $O(kn^{1+1/k})$ size can be computed in {\em StreamSort} model
in $O(k)$ passes total and with $O(\log n)$ bits of working memory only.
Furthermore, each {\em Stream} pass in this algorithm
spends just $O(1)$ time per edge.
\end{enumerate}
We would also like to mention that the algorithms presented in our paper
employ elementary data structures (link lists and arrays). 
The algorithms (and their analysis) presented in this paper are 
complete on their own.

\noindent
{\em Remark}.~ 
Elkin and Zhang \cite{EZ:04} address the problem of computing 
$(1+\epsilon,\beta)$-spanner in streaming environment.
Their algorithm, though sheds some light on the APASP problem in
streaming environment, has little practical relevance. This is because,
the number of passes required, though constant, depend quite heavily on
$\epsilon,\beta$.

%
%
%
%
\section{Preliminaries} 
We assume, like the previous algorithms \cite{FKM+:04,FKM+:05}, that $n$, the 
number of vertices is known in advance and the 
vertices are numbered from 1 to $n$. 

As mentioned in the introduction, our algorithm is basically a careful 
adaptation of the previous static linear time algorithms 
\cite{BS:03,BS-J:03,BKMP:05} in the streaming environment. The central idea of 
these algorithms is clustering which we define below. 
\begin{definition}
A {\bf cluster} is a subset of vertices, and a {\bf clustering} ${\cal C}$, 
is a union of disjoint clusters. Each cluster will have a unique vertex 
which will be called its {\bf center}. 
\label{cluster}
\end{definition}
The uniqueness of the center of a cluster can be used to represent a clustering
${\cal C}$ as an array (of the same label ${\cal C}$) of size $n$ in the 
following way :
${\cal C}(v)$ will denote the center of the cluster containing $v$ unless
when $v$ does not belong to any cluster, in which case ${\cal C}(v)=0$.
We shall say that a cluster $c$ is {\em incident} on or {\em adjacent} to a 
vertex $u$ if there is some vertex $v\in c$ adjacent to $u$. With respect to
a given clustering ${\cal C}$, a vertex $u\in V$ is said to be a 
{\em clustered} vertex if it belongs to some cluster in ${\cal C}$, and 
an {\em unclustered} vertex otherwise.

%
%
The role of clustering to achieve a small size spanner can be described 
intuitively as follows. 
Suppose we can partition the vertices into a small number of disjoint clusters,
and span each of these clusters by a small set ${\cal E}\subseteq E$. 
As a consequence of this clustering, each vertex $u\in V$ has all its 
neighbors grouped in various clusters. Among those edges that are incident 
on $u$ from same cluster, say $c$, selecting just one edge will ensure the 
following property. For each missing edge $(u,v)$ such that $v\in c$, there is 
a path connecting $u$ and $v$ using one of the selected edges and some
edges from ${\cal E}$, and the length of this path is at most one unit more 
than the diameter of the cluster containing $v$. 
(In order to ensure a small bound on the stretch, we need these clusters to 
have very small diameter). 
This simple idea of pruning edges lies at the core of the static algorithm 
of \cite{BS:03,BS-J:03,BKMP:05}, and to materialize it they build a multilevel 
clustering using random sampling. 

%
%
\section{Algorithm for $(2k-1)$-spanners in classical streaming model}
Prior to processing the stream of edges, the algorithm constructs an 
initial $(k+1)$-levels of clusterings $\{{\cal C}_i| 0\le i\le k\}$ for
the empty (without edges) graph as follows. 
\begin{center}
\fbox{\parbox{12cm}{
\centerline{\bf Initializing the $(k+1)$-levels of clusterings}
}}
\fbox{\parbox{12cm}{
\centerline{}
Let $S_0\leftarrow V$,~$S_k=\emptyset$\\
For $0<i<k$,\\
\hspace*{.2in}
$S_i$ contains each element of set $S_i$ independently with prob. $n^{-1/k}$\\
For $0\le i \le k$\\
\hspace*{.2in} ${\cal C}_i\leftarrow \{\{v\}| v\in S_i\}$\\
}}
\end{center}
We introduce two notations at this point.\\
$\ell(v)$~:~ the highest level of the clustering in which $v$ 
is present as a clustered vertex.\\
$\ell_S(v)$~:~ the highest level $i<k$ such that the cluster centered at $v$
is a sampled cluster in ${\cal C}_i$.\\
Note that, in the beginning $\ell_S(v)=\ell(v)$ for all the vertices. However,
as the edges are being processed, the level $\ell(v)$ of a vertex might rise. 

\noindent
We shall now give an overview and intuition of the algorithm. Initially, at 
each level $i<k$, every cluster is a singleton set. 
From viewpoint of clustering, the only change in a cluster during the algorithm
will be that other vertices (from levels lower than the cluster) might 
join it. We shall always use the following convention :
a cluster $c\in {\cal C}_i$ is a sampled cluster if in the beginning of the 
algorithm, the corresponding singleton cluster was a sampled cluster. The 
following assertion will hold throughout.
\begin{center}
${\cal A}$~:~\fbox{\parbox{12cm}{
For each $c\in {\cal C}_{i+1}$, there exists a sampled cluster 
$c'\in{\cal C}_i$ such that $c'\subseteq c$.
}}
\end{center}

Now we describe the way the stream of edges is processed by the algorithm, and 
how the clustering evolves by upward movement of vertices. Each
vertex $u\in V$ waits at its present level $\ell(u)$ for an opportunity
to move to a level higher than $\ell(u)$, and the only opportunity for it to 
move higher is when it receives an edge incident from some sampled cluster in
${\cal C}_{\ell(u)}$. 
We shall explain soon how this tendency of vertices to rise to higher level 
proves crucial to compute a sparse $(2k-1)$-spanner. 
It follows from assertion ${\cal A}$ that a sampled cluster 
$c\in {\cal C}_i$ has some $c'\in {\cal C}_{i+1}$ such that $c\subseteq c'$. 
Whenever $u$ gets such an edge, it hooks itself to the sampled cluster $c$ to 
join (become member of) cluster $c'$ (so $c'$ gets updated accordingly).
In case, $c$ appears as sampled cluster at the next level also, 
the vertex $u$ will join the next level parent as well.
As follows from the sampling involved in building the hierarchy of 
clusterings, only a very few of the clusters
at any level are the sampled clusters. So a vertex will get an 
opportunity to become adjacent to a sampled cluster on very few occasion, and 
until then, it adds edges to the spanner in a frugal manner using the smart 
idea of clustering, as follows. Let the vertex $v$ be member of only unsampled 
cluster at level $\ell(u)$.
Let $c$ be the cluster at level $\ell(v)$ in which $v$ is present.
In this case, the vertex $u$ just adds an edge $(u,v)$ to the spanner if
$c$ was not adjacent to $u$ earlier. Vertex $u$ would keep a list 
storing one edge from each cluster of ${\cal C}_i$ that is 
adjacent to it. Now, in order to determine whether the cluster $c$ was
previously incident on $u$ before the edge $(u,v)$, it suffices to explore the 
entire list of edges incident from various clusters at level $\ell(v)$, which 
could be quite large. 
(Feigenbaum {\em et al.} \cite{FKM+:05} used this brute force search). 
In order to achieve amortized $O(1)$ time, we adopt a buffering approach in 
which we keep a buffer storing the edges at each level temporarily. The vertex 
$v$ will initially add the edge $(u,v)$ to its temporary buffer at level 
$\ell(v)$, and prune this set once there are {\em sufficiently} large number 
of edges using the procedure {\em Prune}($u,i$).

A vertex's tendency to move to higher levels proves crucial to compute a 
sparse $(2k-1)$-spanner in the following way. At lower level, there are a 
large number of clusters, so we can't afford to add edges from a vertex to all 
these clusters. As more and more number of clusters at level $i\ge \ell(u)$ get
adjacent to $u$, one of them might be a sampled cluster. Since a sampled 
cluster is present at higher level too (see assertion ${\cal A}$), 
getting hooked to a sampled cluster would pay $u$ in the sense that
it moves to a higher level where there are fewer clusters. At level $k-1$,
there would be expected $n^{1/k}$ clusters, and once $u$ reaches this level, 
it can afford to add a single edge to each of its neighboring clusters.

Having given an intuitive and informal description of the algorithm 
above, now we shall present the algorithm and the associated data structures 
formally.

\noindent
{\bf Data structure ~}:~ We shall use $k$ arrays ${\cal C}_i, i<k$ to store
clustering at each level. As mentioned earlier ${\cal C}_i(u)$ will
store the center of the cluster in ${\cal C}_i$ storing $u$. In case
$u$ is not clustered at level $i$, ${\cal C}_i(u)$ will store 0. 
Each vertex $u\in V$ keeps lists $Temp(u)$ and ${\cal E}(u)$. The list 
${\cal E}(u)$ will store edges incident on $u$ from unsampled clusters at 
level $\ell(u)$, and $Temp(u)$ will act as a buffer for these edges which we 
shall purge once the number of edges in $Temp(u)$ exceeds the number of edges
in ${\cal E}(u)$.
%
%
%
\begin{center}
\fbox{\parbox{10cm}{
\centerline{\bf Processing an edge $(u,v)$ from the stream}
}}
\fbox{\parbox{10cm}{
\begin{enumerate}
\item {\em Assigning the edge to the endpoint at lower level} \\
  If $\ell(u)>\ell(v)$, then swap $(u,v)$.\\
  $i\leftarrow \ell(u)$~,~$x\leftarrow {\cal C}_i(v)$~,~$h\leftarrow \ell_S(x)$~,
  \item {\em Processing the edge}\\
  {\bf If} $h> i$ \\
\hspace*{.2in}2.1~~ {\bf For} $j=i+1$ to $h$,~{\bf do}\\
\hspace*{.4in}~~~~~~ ${\cal C}_j(u)\leftarrow x$~,\\
\hspace*{.2in}2.2~~ $\ell(u) \leftarrow h$\\
\hspace*{.2in}2.3~~ ${\cal E}_S \leftarrow Temp(u)\cup{\cal E}(u) \cup \{(u,v)\}$\\
\hspace*{.2in}2.4~~ $Temp(u) \leftarrow \emptyset$~,
              ~${\cal E}(v)\leftarrow \emptyset$\\
{\bf Else}\\
\hspace*{.2in}2.5~~ $Temp(u) \leftarrow Temp(u) \cup \{(u,v)\}$.\\
\hspace*{.2in}2.6~~ If $|Temp(u)|~\ge~|{\cal E}(u)|$, then {\em Prune}($u,i$).
\end{enumerate}
}}
\end{center}
%
%
The {\bf If} condition  in step 2 checks whether there is any sampled cluster
containing $v$ at level $\ell(u)$ or higher, and if so, the vertex $u$ 
joins a cluster. Otherwise, the clustering remains unchanged. 
It is easy to observe that the assertion ${\cal A}$ will hold after every edge 
is processed.\\
\centerline{}
{\bf The procedure {\em Prune}($u,i$)} :
The procedure uses a boolean array $A[1..n]$ as a scratch space. The array $A$ 
is initialized to 1. First it scans the list ${\cal E}(u)$ and sets to 1 
entries in $A$ corresponding to clusters in ${\cal C}_i$ neighboring to $u$.
It then scans the edges in the list $Temp(u)$, and eliminates an edge if 
the corresponding cluster was already incident, otherwise it adds it to 
${\cal E}(u)$. Afterwords, we scan the updated list 
${\cal E}(u)$ once to undo the changes made in array $A$ so that $A$ is 
initialized back to its start stage (all entries set to 0).
%
%
\begin{center}
\fbox{\parbox{10cm}{
\centerline{\bf The procedure $Prune(u,i)$}
}}
\fbox{\parbox{10cm}{
\begin{enumerate}
\item {\bf For} each edge $(u,w)\in {\cal E}(u)$, {\bf do}\\ 
\hspace*{.2in}$A[{\cal C}_i(w)]\leftarrow 1$.
\item
{\bf For} each edge $(u,v)\in Temp(u)$, {\bf do}\\
\hspace*{.2in} {\bf if} $A[{\cal C}_i(v)]=0$ ~ and ~ ${\cal C}_i(u)\not={\cal C}_i(v)$,\\
\hspace*{.4in} 2.1~~$A[{\cal C}_i(v)] \leftarrow 1$.\\
\hspace*{.4in} 2.2~~${\cal E}(u) \leftarrow {\cal E}(u) \cup \{(u,v)\}$.\\
\hspace*{.2in} $Temp(u) \leftarrow Temp(u)\backslash (u,v)$.
\end{enumerate} 
}}
\end{center}
%
%
\begin{obs}
For each vertex $u\in V$, $|Temp(u)|\le |{\cal E}(u)|$ except before the 
invocation of $Prune(u)$ when $|Temp(u)|$ exceeds $|{\cal E}(u)|$ by one.
\label{obs1}
\end{obs}

%
%

%
%
\subsection{Analyzing the running time}
It takes $O(1)$ time for processing an edge except when it invokes $Prune()$.
Let us analyze the total time spent in a single call of 
$Prune(u,i)$. It follows from the description of the procedure that the 
total time required by $Prune(u)$ is of the order of 
$|{\cal E}_i(u)|+|Temp_i(u)|$, which by Observation \ref{obs1} is 
$O(|Temp_i(u)|)$. So it suffices to charge $O(1)$ cost to each edge of 
$Temp_i(u)$ to account for the time spent in a call of $Prune(u,i)$. Note that
an edge is processed only once by $Prune(u,i)$ while being a member of 
$Temp_i(v)$. This is because, after $Prune(u,i)$ procedure, either the edge gets
discarded forever or it becomes a member of ${\cal E}_i(u)$.
Hence it suffices to charge $O(1)$ cost to each edge in order to account for the
total computational cost charged to all calls of $Prune()$ during the algorithm. 
Hence total time spent in required for processing the stream of edges is $O(m)$.\\

\noindent
\begin{center}
\fbox{\parbox{15cm}{
\centerline{
Let ${\cal E}^+$ be the set 
$\cup_{i<k,u\in V} (Temp_i(u) \cup {\cal E}_i(u))\cup {\cal E}$ 
at any stage of the algorithm.}}}
\end{center}

\noindent In the following section, we shall prove that : the set 
${\cal E}^+$ at any given moment is a $(2k-1)$-spanner for 
the set of edges appeared in the stream till that moment, and its expected 
size $O(kn^{1+1/k})$. This way, the algorithm can also be viewed as an incremental
algorithm for computing a $(2k-1)$-spanner.
%
%
\section{The stretch and the size of the spanner computed by the algorithm}
%
%
\subsection{Analysis of the stretch of the spanner}
First we state an important Lemma.
\begin{lemma}
Let $c$ be any cluster in ${\cal C}_i$. Each vertex $v\in c$ is 
connected to its center through at most $i$ edges from ${\cal E}$.
\label{radius}
\end{lemma}
\begin{proof}
The proof is based on induction on $i$ and the number of edges of the stream 
seen so far. Let $x$ be the center of the cluster $c$. If $c$ is a singleton 
cluster, there is nothing to prove, so assuming otherwise, let $u\not=x$ be a 
vertex which belongs to $c$. Now observe the process by which $u$ joined the 
cluster $c$. The vertex $u$ became member of $c$ only in the situation where
an edge $(u,v)$ appeared in the stream with
vertex $v$ being a member of some sampled cluster $c'$ in ${\cal C}_{i-1}$.
The assertion ${\cal A}$ implies that, $c'$ is a subset of $c$ and so
has $x$ as its center. Now applying inductive assertion, there is a path
$\subseteq {\cal E}$ between $v$ and $x$ with length $i-1$. This path
concatenated with the edge $(u,v)$ (also in ${\cal E}$), is
a path $\subseteq {\cal E}$ between $u$ and $x$ of length at most $i$.\\
\end{proof}

The streaming algorithm processes each edge of the stream and discards
a dispensable edge only through the procedure $Prune()$.
In order to prove that ${\cal E}^+$ is a $(2k-1)$-spanner,
we shall show that for each edge $(u,v)$ discarded
by the algorithm, there is a path in ${\cal E}^+$ of length at most 
$(2k-1)$ that connects $u$ and $v$. Without loss of generality, assume that
the edge $(u,v)$ got discarded during $Prune(u,i)$.
Now the edge $(u,v)$ could be discarded only if we had already selected 
some other edge $(u,w)$ in ${\cal E}_i(u)$ incident from the same cluster in
${\cal C}_i$ to which $v$ belongs. Lemma \ref{radius} implies that the center 
of each cluster in ${\cal C}_{i}$ is connected to its members through a path 
in ${\cal E}$ with length at most $i$. Hence $v$ and $w$, being the members of 
the same cluster, are connected by a path in ${\cal E}$ with length at most 
$2i$. This path concatenated with the edge $(u,w)\in {\cal E}(u)$, is a 
path in ${\cal E}^+$ between $u$ and $v$ with length at most $2i+1$, 
which is at most $2k-1$ since $i<k$ always. Hence we can conclude that
${\cal E}^+$ at any moment is a $(2k-1)$-spanner for the the set of edges 
appeared in the stream till that moment.
%
%
%
\subsection{Analyzing the size of the spanner}
In the algorithm, a vertex $u$ contributes edges to ${\cal E}$ only when its
level $\ell(u)$ increases. So $|{\cal E}|\le n(k-1)$.
Let us count the expected number of edges in ${\cal E}_i(u)$ and $Temp_i(u)$.
It follows from Observation \ref{obs1} that the number of edges in $Temp_i(u)$ is
at most $|{\cal E}_i(u)|+1$. So it suffices to bound the number of edges in ${\cal E}_i(u)$.

First we would like to make an observation. When an edge $(u,v)$ appears in the stream
with $\ell(u)\le \ell(v)$ and let vertex $v$ does not belong to a sampled 
cluster at any level from $\ell(u)$ onwards. This edges makes $u$ 
adjacent to the cluster containing $v$ at level $\ell(v)$. Note from the 
algorithm that although the vertex $v$ is clustered from every level 
$\ell(u)$ to $\ell(v)$, it is only the cluster at level $\ell(v)$ which gets 
adjacent to $u$ by edge $(u,v)$. So the sets $\{{\cal E}_i(u)\}$ are disjoint always.
It also follows from the procedure $Prune()$ that ${\cal E}_i(u)$ stores one edge per cluster at level $i$ that gets adjacent to $u$. 

We shall give a bound on the expected size of $|{\cal E}_i(u)|$.
For any arbitrary but fixed stream of edges, let 
$\langle c_1,c_2,\ldots\rangle$ be the clusters at level $i$ arranged in the 
chronological order of their getting incident on to $u$. When a cluster 
from ${\cal C}_i$ gets adjacent to $u$ and the cluster is a sampled cluster, 
the vertex $u$ will hook onto that cluster and move to the next level. 
It follows from the algorithm that from this time onwards, $u$ won't add any 
edge to $Temp_i(u)$ or ${\cal E}_i(u)$. So an edge incident from $c_j$
will be selected in ${\cal E}_i(u)$ if none of $c_1,...,c_{j-1}$ were
a sampled cluster. From the sampling of clusters done in the beginning of the 
algorithm, it follows that each cluster at level $i$ is a sampled cluster 
independently with probability $p=n^{-1/k}$. So an edge incident from 
$c_j$ on $u$ will e added to ${\cal E}_i(u)$ with probability $(1-n^{-1/k})^j$.
Hence the expected number of edges in ${\cal E}_i(u)$ is 
\[
	\sum_{1\le j}(1-n^{-1/k})^{j-1}  \le  n^{1/k}
\]
Since there are $n$ vertices, it follows that the expected size of the
spanner computed by the streaming algorithm will be $O(kn^{1+1/k})$. 
Note that it could be that vertex $u$ moves to level higher than $i$
even when it gets adjacent to some sampled cluster at some level $>i$.
But that would only decrease the number of edges contributed as analyzed above.

\begin{theorem}
Given any $k\in \NN$, a $(2k-1)$-spanner of expected size 
$O(\min(m,kn^{1+1/k}))$ for an unweighted graph can be computed in streaming 
model in one pass with amortized constant processing time per edge. The working memory required is $O(kn^{1+1/k})$.
\label{unweighted}
\end{theorem}

Now we shall show that the algorithm for classical streaming model
described above will work for weighted graphs as well if the  edges
appear in the increasing order of edge weights.
 
We shall employ the following observation which follows from the
procedure $Prune()$. 
\begin{obs}
Consider any vertex $u$, $c\in {\cal C}_i,i<k$ and the period during which
$\ell(u)\le i$. Among all the edges in the stream that get incident on $u$ from
$c$ in this period, the edge that appears first in the stream is surely 
present in the spanner.
\label{obs-22}
\end{obs}
\begin{proof}
Let $(u,v), v\in c$ be the first edge incident on $u$ from $c$ during the
period $\ell(u)\le i$. It will be added to $Temp_i(u)$ initially like any other
edge. When $Prune(u,i)$ is invoked in near future, and the edge $(u,v)$ is
processed, it is clear that $A[{\cal C}_i(v)]=0$ since by definition 
there was no edge prior to $(u,v)$ which is incident on $u$ from $c$. Hence
$(u,v)$ gets added to ${\cal E}_i(u)$ and subsequently to the spanner.
\end{proof}

Along similar lines, we can infer the following observation.
\begin{obs}
Consider any cluster $c\in {\cal C}_i$, and let $v$ be a vertex present in
$c$. For the period $\ell(v)=i$, let $E_v$ be the edges that gets incident on
$v$ from vertices lying at level $\le i$. All the edges lying on the
path from $v$ to the center of $c$ appeared before any edge in the set $E_v$.
\label{obs-33}
\end{obs}

Let the edges in the stream appear in the non decreasing order of their 
weights. Let our single pass algorithm (designed for unweighted graph)
processes this stream ignoring the edge weights. We shall show that
the spanner computed will also be a $(2k-1)$-spanner of the original
graph with weighted edges. Let $(u,w)$ be an edge discarded by the algorithm.
and let us suppose it got discarded during $Prune(u,i)$, for some $i<k$.
Let $w\in c\in {\cal C}_i$, it follows from Observation \ref{obs-22}
that there is some edge, say $(u,v),v\in c$ that appeared before $(u,w)$
in the stream and got added to the spanner. From the arguments used in the 
proof of Lemma \ref{radius}, it follows that $v$ and $w$ were connected
by a path of at most $2i$ edges from set ${\cal E}$. All these edges and
the edge $(u,v)$ form a path in the spanner of length at most $2i+1$. Using
Observation \ref{obs-22} and \ref{obs-33}, it also follows that all these 
edges appeared before the edges $(u,w)$ in the stream. Hence each of them is
at most as heavy as $(u,w)$ since the edges appeared in the stream in the
nondecreasing order of their weights. 
So there is a path between $u$ and $w$ in the spanner
consisting of at most $2i+1$ edges each one being at most as heavy as $(u,w)$.
Hence the spanner is indeed a $(2k-1)$-spanner.

Thus we can conclude that our single pass streaming algorithm originally
designed for unweighted graphs will also compute a $(2k-1)$-spanner
for weighted graph provided the edges appear in nondecreasing order of
their weights. So an algorithm for computing a $(2k-1)$-spanner in 
{\em StreamSort} model would be as follows.
\begin{enumerate}
\item First run a {\em sort} pass on the input stream $I$ which will produce
an output stream $O$ where edges appear in the nondecreasing order of their
weights.
\item Execute our single pass algorithm of earlier section (originally designed
for unweighted graphs) on the stream $O$ ignoring the weights.
\end{enumerate}

\begin{theorem}
Given any $k\in \NN$, a $(2k-1)$-spanner of expected size 
$O(\min(m,kn^{1+1/k}))$ for weighted graph can be computed in StreamSort 
model with one sort pass followed by one stream pass and it requires amortized 
constant processing time per edge during the stream pass and the working memory
required is $O(kn^{1+1/k})$.
\label{weighted}
\end{theorem}

In the following section we shall describe an algorithm for computing
$(2k-1)$-spanner in StreamSort model which will require $O(\log n)$ bits of 
working memory and perform $O(k)$ passes only. 
%
\section{Algorithm for $(2k-1)$-spanners in {\em StreamSort} model}
%

We shall now present an algorithm for computing a $(2k-1)$-spanner in 
{\em StreamSort} model. The algorithm works for weighted graphs as well and 
will require just $O(\log n)$ bits of working memory and $O(k)$ alternating 
passes of Streaming and Sorting.

The algorithm can be viewed as a streaming version of the static RAM algorithm
for computing $(2k-1)$-spanner given by \cite{BS:03}. We provide a brief 
overview of the algorithm below.
The algorithm executes $k$ iterations. Each iteration begins with
a partially built spanner $E_S$, a subset of edges $E'$ for which decision of
including them into spanner has yet to be made, a subset 
$V'\subset V$ such that end point of each edge in $E'$ is present in $V'$. 
In addition, $i$th iteration begins with a clustering ${\cal C}_{i-1}$ which 
partitions $V'$ into disjoint clusters such that each edge in $E'$ is an 
inter-cluster edge. The clustering ${\cal C}_{i-1}$ has the following crucial 
property.\\
{\cal P} : For each edge $(u,v)\in E'$, there is a path from $u$ to the center 
of its cluster in ${\cal C}_{i-1}$ with $i-1$ edges each of weight not more 
than that of $(u,v)$.
The first iteration begins with $V'=V, E'=E, E_S=\emptyset, 
C_{0}= \{\{v\} | v\in V\}$.\\

\noindent
Execution of $i$th iteration selects each cluster from ${\cal C}_{i-1}$
independently with probability $n^{-1/k}$. This sampling forms
the basis of defining the clustering for $i$th iteration.
Namely, ${\cal C}_i$ consists of the clusters sampled in $i$th iteration 
with every vertex not belonging to any sampled cluster joining its nearest 
neighboring sampled cluster (if any). In addition to it, processing of 
each vertex in $V'$ contributes some edges to spanner and discards a few in the
$i$th iteration. 
We shall describe the exact description of the $i$th iteration 
and its execution in {\em StreamSort} model soon. But before that, we need to 
proprocess the initial stream of edges, and introduce a few key ideas which 
lead to execution of $i$th iteration in StreamSort model in $O(1)$ passes.

\subsection{Augmenting the initial edge stream, and two sorting primitives}
Our algorithm will receive just a stream of edges. In order to
execute our algorithm, we will associate some more fields with each edge and
vertex. We do so as a preprocessing phase of the algorithm.
{\bf Preprocessing of initial edge stream : }
We preprocess the initial stream of edges to produce another stream such that
for an edge between $u,v\in V$ in the stream, we introduce two edges denoted
as $(u,v)$ and $(v,u)$ in the output stream. We shall use $(u,v)$ to denote 
the edge associated with vertex $u$ and we shall use $(v,u)$ to denote the edge
associated with vertex $v$.\\
In addition, we augment the data structure of each edge $(u,v)$ with
the following additional fields.  
\begin{itemize}
\item {\em lcenter} and {\em rcenter} 
storing the center of cluster to which $u$ and $v$ belong in present 
clustering. Since the initial clustering is $\{\{x\}|x\in V\}$, 
${\cal C}(u)\leftarrow u$ and ${\cal C}(v)\leftarrow v$. 
\item {\em spanner-edge} :
which is set to 1 if $(u,v)$ is selected as spanner, and set to -1 if it has 
not to be added to spanner, and to 0 if no such decision has been made. 
So initially, this field is set to 0 for each edge.
\item {\em sampled-edge} : which is set to 1 if either of $u$ or $v$ belong
to a sampled cluster during an iteration.
\end{itemize}
\noindent
For each vertex $u\in V'$, we store the following additional variables.
\begin{itemize}
\item
${\cal C}(u)$ : the center of the cluster in present clustering 
                containing $u$. Initially ${\cal C}(u) \leftarrow u$.
\item
{\em sampled}$(u)$ : a boolean variable which is true during an iteration if 
                     $u$ belongs to sampled cluster.
\item
${\cal N}(u)$ : the weight of the edge incident on $u$ from nearest 
neighboring sampled cluster.
\end{itemize}

Main idea is to show that for processing various steps of an iteration,
we need to sort the edges and vertices in a suitable total order  such
that each task of $i$th iteration can be executed by performing
a few {\em Sort} passes and a few {\em Stream} passes,
We shall first introduce two total orders on the set of edges.
\begin{enumerate}
\item $\preceq_{0}$\\
An edge $(x,y)$ precede $(p,q)$ in  $\preceq_{0}$ if \\
$\min(x,y)<\min(p,q)$ or $\min(x,y)=\min(p,q)$ and $\max(x,y)<\max(p,q)$.\\

\item $\preceq_{({\cal C},{\cal C}')}$\\
Given two clustering ${\cal C},{\cal C}'$ on a set of vertices $V'$, we define 
an order $\preceq_{({\cal C},{\cal C}')}$ on the set of vertices $V'$ and 
edges $E'$ as follows.
\begin{itemize}
\item
a vertex $u$ would precede vertex $v$ in the total order 
$\preceq_{({\cal C},{\cal C}')}$ if \\
  ${\cal C}(u) < {\cal C}(v)$ \hspace*{.1in} or  \hspace*{.1in} 
  ${\cal C}(u) = {\cal C}(v)$ and ${\cal C'}(u) < {\cal C'}(v)$ \\
  We break the tie, that is, ${\cal C}(u) = {\cal C}(v)$ and 
  ${\cal C'}(u) = {\cal C'}(v)$
  by comparing the labels $u$ and $v$.
\item
an edge $(u,v)$ would precede another edge $(x,y)$ in the order 
$\preceq_{({\cal C},{\cal C}')}$ if \\
  ${\cal C}(u) < {\cal C}(x)$ \hspace*{.1in}  or \hspace*{.1in}
  ${\cal C}(u) = {\cal C}(x)$ and ${{\cal C}'}(v) < {\cal C}'(y)$\\
  We break the tie, that is, 
  ${\cal C}(u) = {\cal C}(x)$ and ${{\cal C}'}(v) = {{\cal C}'}(y)$,
  by resorting to lexicographic comparison of $(u,v)$ and $(x,y)$.
\item a vertex $u$ precede an edge $(x,y)$ in the order 
$\preceq_{({\cal C},{\cal C}')}$ if ${\cal C}(u)\le {\cal C}(x)$.
\end{itemize}
\end{enumerate}
\begin{lemma}
Suppose we want to arrange all the edges so that if there is an edge
between two vertices $u$ and $v$, then its two occurrences $(u,v)$ and
$(v,u)$ occur together. This goal can be achieved by a sorting
according to the order $\preceq_{0}$.
\end{lemma}

We now state the following Lemma which would highlight the importance of 
arranging edges according to the order $\preceq_{({\cal C},{\cal C}')}$.

\begin{lemma}
If the list of edges $E'$ is arranged according to the order 
$\preceq_{({\cal C},{\cal C}')}$, then 
for any two clusters $c\in {\cal C}, c'\in {\cal C}'$, \\
(i)~ the set of edges $\{(u,v) | u \in c\}$, i.e. the edges emanating from 
    the cluster $c$ appear as a sub-list, say $L_c$.\\
(ii)~the set of edges $E'(c,c')$ appear as a sub-list within the 
     sub-list $L_c$.\\
\end{lemma}
\begin{corollary}
If either ${\cal C}$ of ${\cal C}'$ is the clustering $\{\{u\} | u\in V\}$, 
then in the total order $\preceq_{({\cal C},{\cal C}')}$, all edges incident 
on a vertex $u$ appear together as a sub-list and immediately succeed 
the vertex $u$.
\label{cor-2}
\end{corollary}
\noindent
\subsection{Algorithm for $(2k-1)$-spanner in {\em StreamSort} model}
{\bf Algorithm :} \\
As mentioned earlier, the algorithm will execute $k-1$ iterations. The $i$th 
iteration will begin with a tuple $(V',E',E_S,{\cal C}_{i-1})$, where $E_S$ is 
a partially built spanner, $E' \subset E$ consists of those edges for which 
decision of selecting into spanner (or discarding) has not been made yet. 
Moreover, each endpoint of an edge in $E'$ is present in $V'$ and the 
clustering ${\cal C}_{i-1}$ partitions $V'$ into disjoint cluster such that
each edge in $E'$ is an inter cluster edge and the property ${\cal P}_{i-1}$ 
is satisfied :\\

Our algorithm does not do any processing on the edges of $E_S$ and basically
processes only $E'$ and $V'$ in the stream. The various fields of the data 
structures associated with $E'$ and $V'$ store the following information in 
the beginning of $i$th iteration -- the fields $lcenter$ and $rcenter$ of each 
edge $(u,v)\in E'$ store ${\cal C}_{i-1}(u)$ and ${\cal C}_{i-1}(v)$ 
respectively. The {\em sampled-edge} field of each edge is reset, and 
{\em sampled} field of each vertex is also reset. ${\cal N}(v)$ of each vertex 
stores $\infty$.

We now present the four basic tasks of the $i$th iteration for computing a 
$(2k-1)$-spanner and their execution in {\em StreamSort} model as follows.
\begin{enumerate}
\item
{\underline{Forming a sample of clusters}} : \\
{\em Sample each cluster from ${\cal C}_{i-1}$ independently with probability
$n^{-1/k}$. However, if $i=k-1$, then sample no cluster} \\

\noindent
Execution in {\em StreamSort}  model ~:~
Perform a sorting pass on the stream of vertices $V'$ and edges $E'$ according 
to the order $\preceq_{({\cal C}_{i-1},{\cal C}_0)}$. Consequently, the 
vertices (and their edges) belonging to same cluster in ${\cal C}_{i-1}$ appear
together in the stream. We make a {\em Stream} pass on this stream and do
the following. We pick each cluster independently with 
probability $n^{-1/k}$ and set the field {\em sampled} of the vertices 
of the sampled clusters accordingly, and also set the field 
{\em sampled-edge} of each edge emanating from them. 
\item
{\underline{Finding nearest neighboring sampled clusters for vertices}} :\\
{\em For each vertex not belonging to any sampled cluster, if it is adjacent to
one or more sampled cluster, compute the least weighted edge incident from the 
nearest sampled cluster; let ${\cal N}(v)$ stores the weight of the edge.}\\

\noindent
Execution in {\em StreamSort}  model ~:~
We sort the stream according to $\preceq_0$ so that for an edge between $u$ and
$v$, the two occurrences $(u,v)$ and $(v,u)$ appear together. We make a {\em 
Stream} pass and if {\em sampled-edge}$(u,v)$ is set to 1, then we set
{\em sampled-edge}$(v,u)$ to 1 as well.
After this, we sort the stream according to the order 
$\preceq_{({\cal C}_0, {\cal C}_{0})}$. As a result, we can observe 
the following. All edges incident on a vertex $v$ 
appear contiguously in the stream. We process each vertex $v\in V$ in this
stream as follows. If $v$ is not sampled, then we select the least weighted 
{\em sampled-edge} incident on it. If $(v,x)$ is such an edge then we set 
${\cal C}(v) \leftarrow rcenter(v,x)$ (so $v$ gets assigned to the cluster
containing $x$ in ${\cal C}_i$), set {\em spanner-edge}$(v,x)$ to 1 and 
let ${\cal N}(v)$ store weight of the edge $(v,x)$. However, in case, $v$ is 
not adjacent to any marked edge, we set ${\cal N}(v)$ to $\infty$.

\item
{\em Adding edges to the spanner} :\\
Each vertex $v$ not belonging to any sampled cluster does the following :
For each cluster $c\in {\cal C}_{i-1}$, incident on $v$ in the clustering with 
edge of weight less than that of ${\cal N}(v)$, we select the least 
weight edge from $E'(v,c)$ and mark it as a spanner edge.\\

\noindent
Execution in {\em StreamSort}  model ~:~
We perform a Sort pass on the stream according to the order 
$\preceq_{({\cal C}_0, {\cal C}_{i-1})}$ so that all the edges incident
on a vertex from same cluster in the clustering ${\cal C}_{i-1}$ appear
contiguous and a vertex precedes immediately all the edges incident on it.
We process a vertex $v$ in the stream as follows.  
For each cluster incident on $v$ with edge of weight less than
${\cal N}(v)$, we mark least weight edge incident on $v$ from that cluster as 
spanner-edge and mark others as non-spanner edge. 

We make a Sort pass over the stream of edges so that both the occurrences of
an edge are together and then delete both of them if any of them has
{\em spanner-edge} field set to -1. 

\item 
{\em Defining the clustering ${\cal C}_i$} :\\
Keep only those vertices which belong to sampled cluster or were adjacent to
sampled cluster.\\

\noindent
Execution in {\em StreamSort } ~:~ 
We make a {\em Stream} pass and delete all those vertices $v$ for which
$sampled(v)=0$ and ${\cal N}(v)=\infty$.
If a vertex $u$ belonged to a sampled cluster, then it continues to belong to 
same cluster. If it did not belong then unless it is deleted, it was adjacent 
to some sampled cluster and ${cal C}(u)$ was set to the center of new cluster
in the second step. This defines a clustering ${\cal C}_i$ for all the vertices
among $V'$ which survived $i$th iteration. We need to set the $lcenter$ and 
$rcenter$ of each edge now according to the new clustering ${\cal C}_i$. We do
so as follows. We make a {\em Sort} pass on the edges $E'$ and vertices $V'$ 
according to the order $\preceq_{({\cal C}_0,{\cal C}_0)}$. Consequently all 
edges incident on a vertex $v$ will appear together. We assign 
$lcenter(v,w)$ of each edge to ${\cal C}(v)$ and reset
$rcenter(v,w)$. We make a {\em Sort} pass according to $\preceq_0$ so that both
the occurrences of an edge appear together. We then perform a {\em Stream} pass
and for each pair of edges $(u,v)$ and $(v,u)$ that appear consecutive now,
we set $rcenter(u,v)\leftarrow lcenter(v,u)$ and 
$rcenter(v,u)\leftarrow lcenter(u,v)$.

\end{enumerate}
It is obvious that each step of $i$th iteration is executed in {\em StreamSort}
model using a constant number of {\em Stream} passes and {\em Sort} passes.
Since the algorithm is a streaming version on the static RAM algorithm,
its correctness follows from the correctness of the latter. However, for
sake of completeness, we shall now provide an overview of the correctness of 
the algorithm.

A simple inductive argument can be given to show that ${\cal P}_i$ holds
at the end of $i$th iteration. And on this basis, it follows that for any
edge $(u,v)$ that we delete from $E'$, there is a path in the spanner $E_S$ 
with at most $(2i-1)$-edges joining $u$ and $v$. So at the end of the 
algorithm, the set $E_S$ will indeed be a $(2k-1)$-spanner. Also note that
the number of clusters incident on a vertex with weight less than
nearest neighboring sampled cluster in ${\cal C}_{i-1}$ is a geometric random 
variable with mean $n^{1/k}$. Hence expected number of spanner edges 
contributed by a vertex in an iteration is $O(n^{1/k})$. Since there are 
$k-1$ iterations, the expected size of the spanner computed by the algorithm
is $O(kn^{1+1/k})$.

\begin{theorem}
Given any $k\in \NN$, a $(2k-1)$-spanner of expected size 
$O(\min(m,kn^{1+1/k}))$ for weighted graph can be computed in StreamSort 
model with $O(\log n)$ bits of working memory and $O(k)$ sort passes and 
stream passes. Furthermore, it requires constant processing time per edge 
during each stream pass.
\label{weighted-final}
\end{theorem}

%
%
\section{Conclusion and open problems}
We presented single pass algorithm for computing a $(2k-1)$-spanner of
expected $O(kn^{1+1/k})$ size with $O(m)$ processing time for the entire 
stream (amortized constant processing time per edge).
We also showed that in the {\em StreamSort} model, the algorithm can be
extended for weighted graph as well and would require one {\em sort} pass
followed by a {\em stream} pass. However, the working memory in both these
algorithm is of the order of size of spanner, which though optimal for
classical streaming model, is very large. We then provide an algorithm 
for computing spanner in StreamSort model with $O(\log n)$ working memory
and $O(k)$ passes. It can be seen that these two algorithm achieve optimal 
or near optimal performance in all aspects -  number of passes, amortized 
processing time per edge, working memory size in both models.
One aspect which is not truly optimal is the expected size of the spanner 
which is away from the conjectured lower bound by 
a factor of $k$ at most. An important open question is :
Can we get rid of multiplicative factor $k$ from the the size $O(kn^{1+1/k})$
of $(2k-1)$-spanner computed in streaming model? Note that this factor is 
present in case of the static randomized algorithm as well. So either 
a more careful and involved analysis of randomized algorithm would
be required or some fundamentally new approach should be pursued to
answer this question.
\bibliographystyle{abbrv}
\bibliography{bible}

\end{document}